\documentclass[11pt]{article}
\usepackage{amssymb}
\usepackage{amsmath}
\usepackage{blois2002,epsfig}
\usepackage{blois2002}

\def\be{\begin{equation}}
\def\ee{\end{equation}}
\def\bea{\begin{eqnarray}}
\def\eea{\end{eqnarray}}

\input{tcilatex}

\begin{document}

\title{CP VIOLATION: \ THE PAST AS PROLOGUE}
\author{ L. WOLFENSTEIN }
\maketitle

\begin{center}
Carnegie Mellon University

{Department of Physics}\\[0pt]
{5000 Forbes Avenue}\\[0pt]
{Pittsburgh, PA 15213\medskip \smallskip \smallskip }

ABSTRACT\smallskip \smallskip
\end{center}

{CP violation is now measured by three numbers: $\epsilon $ and }${\epsilon }
${$^{\prime }$ for the $K^{0}$ system and $sin2\beta $ for $B^{0}$. A future
measurement of the analogue of $\epsilon ^{\prime }$ for the B system would
end any possibility of a Superweaklike Theory. Future frontiers in CP
violation are briefly discussed.} \newpage

\section{$\protect\epsilon $ and $\protect\epsilon ^{\prime }$ for the $%
K^{0} $ System}

\label{subsec:prod}

C,\ P, and T have played a major role in particle physics over the last
fifty years. \ I like to say that these were theoretical discoveries, not
postulates. \ The starting point was assumed Hamiltonians, in particular
QED, which were found to have these symmetries. \ When Fermi invented the
weak interaction in 1933, he modeled it after QED and so it had these
symmetries, although this was not emphasized for a long time.

In the 1950's it was pointed out that it was easy to invent interactions
that violated these symmetries. \ However, every local relativistic
interaction was found to be invariant under C times P times T. \ Thus, when
P violation was found in 1957, the V-A interaction that explained it
violated P and C but left CP and T invariant.

Then in 1964, CP violation was discovered in the decay $K_{L}\longrightarrow
2\pi .$ \ Thirty-six years later the only CP or T violation known was in the 
$K^{0}$ systems. \ Now we have the first observation of CP violation in the $%
B^{0}$ system.

The first observation could be explained as due to CP violation in $K^{0}-%
\overline{K^{0}}$ \ mixing. \ If we set $\overline{K^{0}}=CP$ times $K^{0}$
then the CP eigenstates are

\begin{eqnarray*}
K_{+} &=&\left( K^{0}+\overline{K^{0}}\right) /\sqrt{2} \\
K_{-} &=&\left( K^{0}-\overline{K^{0}}\right) /\sqrt{2}
\end{eqnarray*}

Then CP violation can result from a small term $m^{\prime }$ \ in the mass
matrix; in the $K_{+}-K_{-}$ \ representation

\begin{equation}
M=%
\begin{pmatrix}
M_{1} & im^{\prime } \\ 
-im^{\prime } & M_{2}%
\end{pmatrix}
\tag{1}
\end{equation}%
\bigskip

where the phase factor $i$ is required by CPT invariance. \ Then the
eigenstates are (to lowest order in $\epsilon $)

\begin{eqnarray}
K_{S} &=&K_{+}+\epsilon K_{-}  \notag \\
K_{L} &=&K_{-}+\epsilon K_{+}  \notag \\
\epsilon &=&\frac{-im^{\prime }}{\left( M_{1}-M_{2}\right) -i\left( \Gamma
_{1}-\Gamma _{2}\right) /2}  \TCItag{2}
\end{eqnarray}

The observed decay $K_{L}\longrightarrow \pi ^{+}\pi ^{-}$ is then due to
the component $\epsilon K_{+}$ in $K_{L}:$

\begin{equation}
\eta _{+-}=\frac{A\left( K_{L}\longrightarrow \pi ^{+}\pi ^{-}\right) }{%
A\left( K_{S}\longrightarrow \pi ^{+}\pi ^{-}\right) }=\epsilon  \tag{3}
\end{equation}

with a magnitude $2\times 10^{-3}.\bigskip $

Twenty five years later, the magnitude of $\epsilon $ was the only quantity
measuring CP violation. \ Another measurement, the lepton asymmetry in $%
K_{L} $ decay, determined $\func{Re}\ \epsilon $ \ and there were
independent measurements of $\varphi _{+-},$ the phase of $\eta _{+-},$
essentially equal to the phase of $\epsilon .$ \ However, all these
experiments did was to confirm CPT invariance, which already predicts very
accurately from Eq. (2) the phase of $\epsilon $

\begin{equation*}
\varphi _{\epsilon }\approx \varphi _{+-}=\tan ^{-1}\left( \triangle
M/\triangle \Gamma \right)
\end{equation*}

Since $\triangle M$ and $\triangle \Gamma $ have nothing to do with CP
violation, the value of $\varphi _{\epsilon }$ in no way is a \underline{%
measure of CP violation}. \ As long as the only CP violation observed can be
attributed to mixing and CPT invariance holds, all CP violation can be
attributed to one real parameter, $m^{\prime }$ \ in the mixing matrix. \
Sometimes a fundamental distinction is made between CP violation just due to
mixing, $\func{Re}\ \epsilon ,$ and CP violation involving interference
between mixing and decay . \ However, since the phase of $\epsilon $ has
nothing to do with the nature of CP violation, this distinction can be
misleading.

After the discovery of CP violation it was pointed out that it could be
explained by a very weak new interaction that changed strangness by 2 units $%
\left( \triangle S=2\right) $ and violated CP. \ Such a four-fermion
interaction could have a coupling $G_{sw}$ equal to $10^{-10}$ to $10^{-11}$
times the Fermi constant $G_{F}.$ \ This came to be called the superweak
model.

A major development in weak interaction theory was the spontaneously broken
gauge theory. \ With the discovery of neutral currents in 1973, this became
the standard model of weak interactions. \ However, the theory had the same
CP invariance as the (V-A) theory.

Various possible methods of extending the theory to allow CP violation were
proposed. \ One of these was in one paragraph of a paper in the "Progress of
Theoretical Physics" that few people read. \ It said that if there were six
quarks instead of four (although the fourth quark $c$ had not yet been
detected) then it was possible to have CP violation. \ With the discovery of
the $b$ quark this became the standard Kobayashi-Maskawa Theory of CP
violation.

To disprove the superweak model it was necessary to detect CP violation in
the decay amplitude. \ This is done by looking for a difference between the
CP violation for the final $\pi ^{0}\pi ^{0}$ state and that for $\pi
^{+}\pi ^{-}.$ \ The parameter $\epsilon ^{\prime }$ is defined by

\begin{eqnarray}
\eta _{+-} &=&\epsilon +\epsilon ^{\prime }  \notag \\
\eta _{00} &=&\epsilon -2\epsilon ^{\prime }  \notag \\
\eta _{00} &\equiv &\frac{A\left( K_{L}\longrightarrow \pi ^{0}\pi
^{0}\right) }{A\left( K_{S}\longrightarrow \pi ^{0}\pi ^{0}\right) }
\end{eqnarray}

After 35 years of experiments at Fermilab and CERN, results have converged
on a definitive non-zero result

\begin{equation}
\func{Re}\ \left( \epsilon ^{\prime }/\epsilon \right) =\left( 18\pm
3\right) \times 10^{-4}
\end{equation}

While in principle $\epsilon $ \ could be attributed to CP violation in
decay or mixing, $\epsilon ^{\prime }$ \ is unambiguously a measure of CP
violation in decay, and thus represents the first evidence against
superweak. \ However, the value of $\epsilon ^{\prime }$ \ is only $4\times
10^{-6}$ so that, while it is not inconsistent with the standard model, it
could be explained by some very weak $\left( \sim 10^{-6}G_{F}\right) $ \
new interaction.

\section{$\protect\epsilon $ and $\protect\epsilon ^{\prime }$ for the B
System}

When it was discovered that the B had a relatively long lifetime and decayed
primarily to charm it became apparent that the CKM matrix had a hierarchical
form\ $[1]$. \ Elements $V_{ij}$ could be expanded in even or odd powers of $%
\lambda $ ($\lambda =\sin $ of the Cabibbo angle). \ In particular,

\begin{eqnarray}
V_{cb} &=&A\lambda ^{2}+0\left( \lambda ^{4}\right)  \notag \\
V_{ub} &=&A\lambda ^{3}\left( \rho -i\eta \right) +0\left( \lambda
^{5}\right) =\left\vert V_{ub}\right\vert e^{-i\gamma }  \notag \\
V_{td} &=&A\lambda ^{3}\left( 1-\rho -i\eta \right) +0\left( \lambda
^{5}\right) =\left\vert V_{td}\right\vert e^{-i\beta }
\end{eqnarray}

Measurements of $V_{cb}$ give $A$ \ between .76 and .9. \ The parameter $%
\eta $ is the source of CP violation. \ With this parameterization it became
clear that large CP-violating effects could be found in B decays in contrast
to K decays.

The phase of $V_{td},$ $\beta ,$ was expected to lie between 10 and 35
degrees given the value of $\eta $ required to fit $\epsilon $ in the K
system. \ Because $B-\bar{B}$ mixing was dominated by the box diagram with
virtual $t$ quarks, the mixing matrix $M_{12}$ would be proportional to $%
V_{td}^{2}$ and so have the phase $2\beta .$ \ The time-dependent CP
violation in $B\left( \bar{B}\right) \longrightarrow \Psi K_{s}$ is then
proportional to $\sin 2\beta .$ \ This is the first CP-violating observable
found in the B system. \ The large value $[2]$ $\sin 2\beta =.73\pm .06$ is
consistent with the expectation of the standard model. \ I referred to this $%
[3]$ as $\epsilon _{B}$ because its origin could be entirely CP violation in
mixing and so could be blamed on\ a new superweak $\triangle B=2$ \
interaction with $G\left( \bar{b}\ d\ \bar{b}\ d\right) \sim 10^{-7}G_{F}.$

The next goal in B physics should be the analog of $\epsilon ^{\prime };$
that is, a CP-violating effect that demonstrates CP violation in the $%
\triangle B=1$ decay amplitude. \ In contrast to the value $4\times 10^{-6}$
in the K system, $\epsilon _{B}^{\prime }$ should be of order unity. \ This
measurement should then be the final blow to any superweaklike theory.

The CP-violating phase in the $b\longrightarrow u$ \ transition has a phase $%
\gamma $ relative to $b\longrightarrow c$ transition (which is involved in $%
B\longrightarrow \Psi K_{s}$). Allowed values of $\gamma $ are large$\colon
\left\vert \sin \gamma \right\vert >\frac{1}{2}.$ \ Thus, by measuring the
CP-violating time-dependent asymmetry $A_{i}$ in some decay $%
B\longrightarrow X_{i}$ where $X_{i}$ is a CP-eigenstate involving $%
b\longrightarrow u$ we can define

\begin{equation}
\epsilon _{B_{i}}^{\prime }=A_{i}-\sin 2\beta
\end{equation}

The example most discussed is $X_{1}=\pi ^{+}\pi ^{-}$ ; considering only
the tree approximation for the decay%
\begin{equation}
\epsilon _{B_{1}}^{\prime }=\sin 2\left( \beta +\gamma \right) -\sin 2\beta
\end{equation}

As pointed out by Winstein, $\left[ 4\right] $ this may accidentally equal
zero for $\gamma =\pi /2\ -2\beta .$ \ In fact, one expects a sizeable
penguia contribution to this decay. \ Estimating this $\left[ 5\right] $ one
finds that $\epsilon _{B1}^{\prime }$ is close to zero for $\gamma $ in the
neighborhood of 50$^{\circ }.$ \ On the other hand, for $\gamma \sim $75$%
^{\circ },$ $\left\vert \epsilon _{B_{1}}^{\prime }\right\vert $ would be
greater than 0.5.

Another possibility are experiments that measure $\sin \left( 2\beta +\gamma
\right) ;$ we define

\begin{equation}
\epsilon _{B_{2}}^{\prime }=\sin \left( 2\beta +\gamma \right) -\sin 2\beta
\end{equation}

Unfortunately, for a fair range of expected values of $\gamma ,$ $\epsilon
_{B_{2}}^{\prime }$ is not very different from zero.

\section{Long Term Frontiers for CP Violation}

\bigskip There are important issues about CP violation which will take many
years to study and some of which may be impossible to resolve. \ These are
briefly discussed in this section.

\subsection{\protect\bigskip Precision CKM Physics}

\bigskip We want to know whether all CP violation can be explained by the
CKM matrix or, alternatively, there are some detectable effects from new
physics. \ For this purpose, we seek observables that can be analyzed with
small theoretical errors. \ The first of these is $\sin 2\beta $ and the
second we expect (hopefully from CDF) is

\begin{equation}
\left[ \triangle M_{d}\ /\ \triangle M_{s}\right] ^{1/2}=\left\vert V_{td}\
/\ V_{ts}\right\vert \ /\ \xi
\end{equation}

$\xi $ is a measure of $SU\left( 3\right) $ violation given by a quenched
lattice calculation $\left[ 6\right] $ as $1.15\pm \ .04,$ but more theory
is needed to get an unquenched value. \ These two will define a narrow
region in the $\left( \rho ,\eta \right) $ plane, one that is derived
entirely from $B-\bar{B}$ mixing. \ If there are new physics contributions
to the mixing; for example, due to SUSY particles or extra Higgs bosons,
then these will not be correct values of $\left( \rho ,\eta \right) .$ \
Thus, we want precision determinations based on decays to see if they are
consistent. \ In particular, there are a variety of experiments aimed at
determining $\gamma $ from decays that involve the $b\longrightarrow u$
transition. \ It should be emphasized that the main purpose of such
experiments is \underline{not} to measure $\gamma $, but to show consistency
(or the lack of it) between the value from decay with that derived
indirectly from mixing.

In addition to B decays, it is also important to use rare K decays. \ In
particular, the decay $K_{L}\longrightarrow \pi ^{0}\nu \vec{\nu}$ $\left[ 7%
\right] $ provides an independent determination of $\eta .$

\subsection{Electric Dipole Moments\protect\bigskip}

Considering only the CP violation due to $\eta $ the electric dipole moment
of the neutron $\smallskip d_{n}<10^{-31}\ e-cm$ \ and that of the electron $%
d_{e}$ is much less. \ Thus, any foreseeable non-zero measurement would be a
signal of new physics. \ In the case of $d_{n},$ the result could be blamed
on $\circleddash _{QCD},$ an arbitrary parameter in the standard model. \ A
non-zero $d_{e}$ would be clearly a signal of physics beyond the standard
model.

\subsection{CP Violation in Lepton Mixing}

Lepton-quark symmetry suggests that there should be CP violation in lepton
mixing analogous to CKM. \ Since the mixing is meaningless in the limit of
zero neutrino masses, the mixing shows up only in neutrino phenomena. \ We
now have strong evidence of neutrino mixing from neutrino oscillations. \
The hope for seeing CP violation is a comparison of $\nu _{\mu
}\longrightarrow \nu _{e}$ \ and $\vec{\nu}_{\mu }\longrightarrow \vec{\nu}%
_{e}$ in long baseline experiments. \ A major problem is that any CP
violation is proportional to the element of $V_{e3},$ which may be zero. \ ($%
V_{e3}$ \ is the $\nu _{e}$ component in the state 3 which is separated from
the two others by $\triangle m^{2}\sim 3\times 10^{-3}ev^{2}$). \ Thus, a
determination of $V_{e3}$ must be the first priority of forthcoming
experiments.

There are in fact two other phases (not present in the quark CKM matrix) for
the case of Majorana neutrinos. \ These have an effect on neutrinoless
double beta decay $\left( \beta \ \beta \right) ,$ but they may be very hard
to detect even if $\beta \ \beta $ is observed $\left[ 8\right] .$ \ 

\subsection{Fundamental Origin of CP Violation}

From the point of view of the standard model there is nothing fundamental
about CP invariance. \ It is violated everywhere it can be and it just turns
out, given the gauge theory and the assumed particle content, that the only
CP violation is in the Yukawa couplings of fermions to Higgs bosons. \
Furthermore, although the Yukawa coupling can contain a number of phases, it
turns out that only one phase is physically significant for quarks (and one
for Dirac neutrinos), that given in the CKM matrix.

There is also one other parameter, called $\circleddash _{QCD},$ which must
be set to a very small value less that $10^{-9}$ to fit limits on $d_{n}.$ \
Both to explain the small value of $\circleddash _{QCD}$ and in hope of a
better understanding of CP violation there are proposals that CP \ is a
fundamental symmetry at some high mass scale which is spontaneously broken.
\ In such theories, $\circleddash _{QCD}\ $is calculable and presumably
non-zero at some loop level. \ It is possible that\ $\eta $ is also
calculable and leads to the standard model $\left[ 9\right] .$ \
Alternatively, most CP violation could be due to new physics if $\eta $
turned out to be too small\ $\left[ 10\right] $.

It is hard to know whether questions such as these will ever be answered. \ 

\subsection{Baryon-antibaryon Asymmetry}

A major hope much discussed in this conference is that somehow CP violation
at some high mass scale effective in the early universe may lead to the
predominance of baryons over antibaryons in the present universe. \ It is
clear that this requires some CP violation beyond the standard model. \ Like
so many cosmological problems, we do not know whether there is some
fundamental answer or whether this is just a peculiarity of our universe
associated with some chaotic birth pangs.\bigskip \newpage

\section*{Acknowledgments}

This work has been supported in part by the U.S. Department of Energy work
grant No.

DE-FG 02-91 ER 40682\bigskip .


\section*{References}

\begin{thebibliography}{99}
\bibitem{[1]} L. Wolfenstein, \emph{Phys. Rev. Lett.}, \textbf{51}, 1945
(1983).

\bibitem{[2]} See the talks by Teramoto and Giorgi.

\bibitem{[3]} L. Wolfenstein, {\emph{Nuc. Phys., }B} \textbf{246}, 45 (1984).

\bibitem{[4]} B. Winstein, {\emph{Phys. Rev. Lett.,} }\textbf{68}, 1271
(1992); B. Winstein and L. Wolfenstein,{\emph{\ Rev. Mod. Phys.,}} \textbf{65%
}, 1113 (1993).

\bibitem{[5]} L. Wolfenstein and F. Wu, \emph{Europhysics Lett.}{\emph{,}} 
\textbf{58}, 49 (2002).

\bibitem{[6]} UKQCD Collaboration, {\emph{Phys. Rev., }D} \textbf{64},
094501 and reference therein (2001).

\bibitem{[7]} See the talk by D. Bryman.

\bibitem{[8]} See the talk by S. Petcov.

\bibitem{[9]} A. Nelson, {\emph{Phys. Lett.,} }\textbf{136 B}, 387 (1984), 
\textbf{143 B}, 165 (1984).

\bibitem{[10]} S. Barr, {\emph{Phys. Rev., }D} \textbf{30}, 1805 (1984).
\end{thebibliography}
\end{document}